\documentstyle[epsf,aps,twocolumn]{revtex}

 \def\DoubleZ{{\rm\bf Z}} \def\half{\frac{1}{2}}

\begin{document}

%\draft

\title{Detecting periodicity in experimental data using linear modeling
  techniques} \author{Michael Small\footnote{Corresponding author: Michael
    Small, Department of Mathematics, University of Western Australia,
    Nedlands, WA 6907, Australia. Tel: +61 8 9380 1359; Fax: +61 8 9380
    1028; Email: {\tt watchman@maths.uwa.edu.au}.} \and Kevin Judd}
\address{Centre for
  Applied Dynamics and Optimization\\Department of Mathematics\\University
  of Western Australia}
\date{\today}

\maketitle
\begin{abstract}
  Fourier spectral estimates and, to a lesser extent, the autocorrelation
  function are the primary tools to detect periodicities in experimental
  data in the physical and biological sciences. We propose a new method
  which is more reliable than traditional techniques, and is able to make
  clear identification of periodic behavior when traditional techniques do
  not. This technique is based on an information theoretic reduction of
  linear (autoregressive) models so that only the essential features of an
  autoregressive model are retained. These models we call {\em reduced
    autoregressive models (RARM)}. The essential features of reduced
  autoregressive models include any periodicity present in the data.
  
  We provide theoretical and numerical evidence from both experimental and
  artificial data, to demonstrate that this technique will reliably
  detect periodicities if and only if they are present in the data. There
  are strong information theoretic arguments to support the statement that
  RARM detects periodicities if they are present. Surrogate data
  techniques are used to ensure the converse. Furthermore, our
  calculations demonstrate that RARM is more robust, more accurate, and
  more sensitive, than traditional spectral techniques.
\end{abstract}

\section{Introduction}

Periodic, and nearly periodic, behavior is a common feature of many
biological and physical systems and there exist several widely-used
techniques to estimate the period of a behavior, for example, spectral
estimation \cite{mP89}, autocorrelation \cite{mP89}, spectrographs, band
pass (comb) filters \cite{pB80b} and wavelet transforms \cite{cB98,cD97}.
All of these standard techniques either employ, or are related to, or are
a generalization of, Fourier series.

In this paper we propose an alternative method of detecting periodicity
that is not so closely related to Fourier series. This new technique
applies ideas from information theory to linear autoregressive models of
time series to extract evidence of periods.

The basic principle is the following. Given a time series
$\{y_{t}\}_{t=1}^{N}$ one can propose a linear autoregressive model
$AR(n)$ by
\begin{eqnarray}
\nonumber  y_{t} & = &a_{1}y_{t-1}+a_{2}y_{t-2}+a_{3}y_{t-3}+\ldots\\
&&+a_{n}y_{t-n}+e_{t}\;\; t=n+1,n+2,\ldots,N.
\label{arn}
\end{eqnarray}
where $e_{t}$ are assumed to be independent and identically distributed
random variables, which are interpreted as the modeling
errors~\cite{mP89,hT90}. Under these assumptions the maximum likelihood
estimate of the parameters $a_1,a_2,\ldots,a_n$ can be written in terms of
a covariance function, and are therefore related to the autocorrelation
function and Fourier spectrum. It is common practice to determine the
optimal size~$n$ of the model by using either the Akaike \cite{hA74} or
the Schwarz \cite{gS78} information criteria; this is done to avoid
over-fitting of the time series~\cite{lA95a}. It has recently been
observed that a further optimization of an $AR(n)$ model may be possible
by deleting some of the terms to obtain a model
\begin{eqnarray}
\nonumber  y_t & = &
a_0+a_{1}y_{t-\ell_{1}}+a_{2}y_{t-\ell_{2}}+a_{3}y_{t-\ell_{3}}+\ldots \\
&& +a_{k}y_{t-\ell_{k}}+e_{t},\label{rarm}
\end{eqnarray}
where,
\[
1\leq\ell_{1}<\ell_{2}<\ell_{3}<\ldots<\ell_{k}\leq n
\]
for $\ell_{i}\in
\DoubleZ^{+}\;\;i=1,2,3,\ldots,k$. The hope is to obtain a model that fits the time series equally well, but
has far fewer parameters. Profound theoretical arguments, which are a
codification of Occam's razor, imply that if a {\em reduced autoregressive
  model} (RARM) is suitably optimized, then it is superior to an
equivalent autoregressive model $AR(n)$.  The key principle of this paper
is that if one has an optimized RARM, that is the RARM that has been
reduced to only the essential terms, then the parameters $\ell_{1},
\ell_{2}, \ell_{3},\ldots, \ell_{k}$, often called {\em{}lags}, provide
information about the periodicity of the time series.

A practical procedure for obtaining an optimal reduced autoregressive
model (RARM) has been described by Judd and Mees~\cite{kJ95a}. This
procedure was introduced in the more general context of nonlinear
modeling, but in the following section we describe briefly the underlying
theory in the context of RARM.

The major part of this paper is aimed at presenting evidence that
examining the lags of an optimal RARM provides a more robust and accurate
means of detecting periods in time series than traditional spectral
techniques. That is, the proposed technique unambiguously identifies
periodicities even when spectral methods fail to do so, and furthermore,
does not falsely suggest the presence of periods when none are present.
The evidence presented is a combination of theoretical argument and
numerical procedures, which are illustrated with both artificial and
experimental data.

An important numerical procedure that will be used to establish that the
proposed technique does not falsely identify periods is {\em{} surrogate
  data analysis}.  The principle of surrogate data analysis is the
following. From experimental data one generates artificial data that are
``similar'' to the experimental data and satisfy a given hypothesis. One
then calculates a test statistic for each surrogate data set, and hence
obtains an ensemble of statistic values that estimate the distribution of
the test statistic under the assumption that the original data is
consistent with the given hypothesis.  One then compares the statistic
value of the original data with the estimated distribution of the
surrogates. If the data has an atypical statistic value then the
hypothesis will be rejected, otherwise it should be accepted. In this
paper we employ this technique to ensure that RARM procedures do not
spuriously identify periodicities in temporally uncorrelated surrogate
data.

\subsection{Minimum description length}
\label{mdl}
The criteria we use for determining the optimal RARM is the minimum
description length. Occam's razor recommends that the best description of
a phenomenon is the shortest description. This principle can be made
rigorous using information theory, and the principle was independently
developed by Wallace~\cite{vH84} and Rissanen~\cite{jR89}.

Operationally the principle is applied as follows. Suppose you have a time
series $\{y_{t}\}_{t=1}^{N}$ given to a certain fixed accuracy and that
you wish to communicate the data to a colleague. To send the raw data
would require a certain number of bits.  Alternatively, one could build a
predictive model, of the form~(\ref{rarm}) for example, and then send the
model parameters (to some precision), the initial~$\ell_k$ observations,
and the differences between the model's predictions and actual
observations.  Given this information your colleague can reconstruct the
original data.  If the model of the time series is good, then the total
number of bits required for parameters, initial conditions and prediction
errors is less than the number of bits of raw data, because the
differences between the predicted and actual observations are smaller than
the observations.  The total number of bits sent in the second case is
called the {\em{}description length}, and the model that achieves the
minimum description length is the one recommended by the application of
Occam's razor. The dogma is that this model achieves the best prediction
of the data without over-fitting.

In practice it is usually sufficient to estimate the description length of
a model, rather than calculate it in detail. An estimate will usually have
the form
\begin{eqnarray*}
&&\left(\mbox{description length}\right)   \approx  (\mbox{number of
  data})\\
&&\hspace{5mm}\times\log\left(\mbox{sum of squares of prediction errors}\right)\\
&&\hspace{5mm}+\left(\mbox{penalty for number and accuracy of
  parameters}\right).
\end{eqnarray*}

Following Judd and Mees~\cite{kJ95a} the description length of a RARM can
be estimated as follows. Given a time series $\{y_{t}\}_{t=1}^{N}$ define
a set of vectors $\{V_i\}_{i=1}^n$ by
\[
V_0 = (1,1,\ldots,1)^T,
\]
\[
V_1 = (y_n,\ldots,y_{N-1})^T,
\]
\[
V_2 = (y_{n-1},\ldots,y_{N-2})^T,
\]
\[
\vdots
\]
\[
V_j = (y_{n-j+1},\ldots,y_{N-j})^T,
\]
\[
\vdots
\]
\[
V_n = (y_{1},\ldots,y_{N-n})^T,
\]
and define
\[
y = (y_{n+1},\ldots,y_N)^T.
\]
Observe that if the model~(\ref{rarm}) is appropriate for the time series
one can write
\begin{equation}
\label{model}
y =  \sum_{i=1}^k a_i V_{\ell_i}+e
\end{equation}
\[
= V_Ba_B+e,
\] 
where $B=(\ell_1,\ell_2,\ldots,\ell_k)$, $V_B=[V_{\ell_1} | V_{\ell_2} |
\cdots | V_{\ell_k}]$ is a matrix, and $a_B=(a_1,a_2,\ldots,a_k)^T$. The
maximum likelihood estimates of $a_B$, that is, the values that minimize
$e^Te$, are given by
\[
a_B = (V_B^TV_B)^{-1}V_B^Ty.
\]
Now each parameter $a_j$ must be sent to some precision $\delta_j$, that
is, the maximum likelihood estimate of $a_j$ is ``rounded-off'' by an
amount $\delta_j$. It can be shown~\cite{kJ95a} that the optimal
precisions $\delta=(\delta_1,\delta_2,\ldots,\delta_k)$, that is, the
round-off for each $a_j$ that gives the minimum description length,
satisfy
\[(Q\delta)_j=1/\delta_j\]
where
\[
Q = \frac{-NV_B^TV_B}{(a_B V_B-y)^T(a_B V_B-y)}.
\]
Consequently, it can be shown~\cite{kJ95a} that the approximate
description length of the RARM~(\ref{rarm}) is
\begin{equation}
\label{mdl2}
\frac{N}{2}\left(1+\ln{\frac{2\pi e^Te}{N}}\right)+\left(\half+\ln\gamma\right)k-\sum_{j=1}^k\ln\delta_j,
\end{equation}
where $\gamma$ is a constant depending on the overall scale of the data.

Armed with this estimate of the description length of a RARM one can
search over all combinations of lags $B = (\ell_1,\ell_2,\ldots,\ell_k)$
to obtain the optimal RARM, however, Judd and Mees~\cite{kJ95a} describe a
fast and efficient method of doing this optimization.

\section{Detecting periodicity using optimal RARM}
\label{rarm-sec}
\label{theory}

A function $f$ is periodic with period $\tau$ if $f(t)=f(t+\tau)$ for all
$t$. A time series (assumed stationary) has an (approximate) periodicity
of period $\tau$ if $y_t\approx y_{t+\tau}$ for all $t$, or, equivalently,
the autocorrelation~$\rho$ has a local maximum at~$\tau$. The reduced
autoregressive model~(\ref{rarm}) predicts the current value of a time
series~$y_t$ as a weighted average of the previous values, that is, at the
time steps $\ell_1$, $\ell_2$, $\ldots$, and $\ell_k$ previous to~$t$. If
a time series has periodic behavior, then the lags $\ell_1, \ell_2,
\ldots, \ell_k$ should be (multiples of) the periods.

We claim that one can detect in time series a periodicity of period
$\leq{}n_{MAX}$ by the following procedure, called the RARM procedure. For
$n=0,1,2,3,\ldots,n_{MAX}$ build optimal reduced autoregressive models of
the form~(\ref{rarm}) using the algorithm described by Judd and
Mees~\cite{kJ95a}. For each model in this sequence calculate its
description length~(\ref{mdl2}) and take as the overall optimal model that
model with the smallest description length.  We claim that if the overall
optimal RARM is non-trivial, then the lags $\ell_1, \ell_2, \ldots,
\ell_k$ should be (multiples of) the periods $\leq{}n_{MAX}$ in the
original time series if the time series is sufficient long.

In order to establish our claim we must demonstrate that
\begin{enumerate}
\item \label{ifstate} {\em if} the time series contains a period {\em
    then} the RARM procedure detects this periodic behavior, and
\item \label{onlyifstate} {\em if} the RARM procedure detects a period
  {\em then} there is periodic behavior in the time series.
\end{enumerate}
In section \ref{forward} we provide a theoretical argument to establish
the forward implication (\ref{ifstate}). In section \ref{reverse} we
discuss an essential procedure for ensuring (\ref{onlyifstate}).

\subsection{Forward implication (\ref{ifstate})}
\label{forward}

The argument to establish the forward implication proceeds as follows.
First, we observe that a period in a time series will (regardless of
whether it is linear or nonlinear) produce a local maximum in the
autocorrelation function~$\rho(\tau)$.  Next it is shown below that, in
the optimization of a RARM of given maximum size~$n$, the criterion for
inclusion of a particular term $a_jy_{t-\ell_j}$ in (\ref{rarm}) is
closely related to the magnitude of the autocorrelation at $\ell_j$,
$\rho(\ell_j)$.  Hence, if $n$ is large enough, the optimal RARM will
include a term corresponding to this periodicity.  Rissanen's minimum
description length criterion guarantees that provided the time series is
sufficiently long this will always be the case and so the RARM procedure
will always detect periods that are present in a time series, provided the
time series is sufficiently long.

The remainder of this section elaborates on the detail of this argument.
A period $\tau$ in a time series $\{y_t\}_{t=1}^N$ of $N$ scalar
measurements is a strong positive correlation between values separated by
$\tau$ time steps, i.e. the autocorrelation
\begin{equation}
\rho(\tau) =  \frac{(y-\overline{y})^T(V_{\tau}-\overline{y})}{\sum_{n+1}^{N}
  (y-\overline{y})^2}
\label{acorrform}
\end{equation} has a local maximum at $\tau$. 
Without loss of generality we may assume that $\overline{y}=0$, and
therefore (\ref{acorrform}) reduces to
\begin{equation}
\label{perform}
 \rho{(\tau)} = \frac{V_\tau^Ty}{y^Ty}.
\end{equation}
Let the set of lags for the optimal RARM of size $k$ be denoted by
$B_k=(\ell_1^{(k)},\ell_2^{(k)},\ldots,\ell_k^{(k)})$. The vector $B_k$
uniquely determines the least squares model
\[
y=\sum_{i=1}^k a^{(k)}_iV_{\ell_i^{(k)}}+e.
\]
Define
\begin{eqnarray}
\nonumber L(\tau) & = & \left|V_\tau^Ty-\sum_{i=1}^{k}a^{(k)}_iV_\tau^TV_{\ell_i^{(k)}}\right|\\
\label{ell_sel} & = & y^Ty\left|\rho(\tau) - \sum_{i=1}^ka^{(k)}_i\rho(\tau-\ell_i^{(k)})\right| 
\end{eqnarray} 
According to the algorithm of Judd and Mees \cite{kJ95a}, given $B_k$ and
$a^{(k)}_B$, the next best term to add to the model has the lag $\tau$
that maximizes $L(\tau)$. However, identity (\ref{ell_sel}) implies that
such a $\tau$ is a local maximum of $\rho(\tau)$.

Rissanen's minimum description length ensures that, for sufficiently large
$N$, ``if there is any machinery behind the data, which restricts the
future observation in a similar manner as the past and which can be
captured by the selected class of parametric functions, then we will find
that machinery'' \cite{jR89}. The argument in the preceding paragraphs
demonstrates that RARM are a sufficiently broad class of parametric
functions to capture ``machinery'' behind the data, including observed
periodicities. Thus, if periodicity is present in the data then RARM
techniques will detect it --- provided $N$ is sufficiently large. This
ensures the forward implication (\ref{ifstate}).

\subsection{Reverse implication (\ref{onlyifstate}): {S}urrogate data techniques}
\label{reverse}
\label{surrrev}
In order to establish that the RARM techniques does not falsely identify a
period when none is present, the numerical procedure of surrogate data
analysis can be used.  The technique of surrogate data was originally
introduced by Theiler and colleagues~\cite{jT92}. They suggest three
surrogate generation techniques to address three different hypotheses
about a time series, but for our purposes we only use Theiler's
algorithm~$0$ surrogates.

In the present case we are interested in whether a time series contains
periodicities, or said in another way, we wish to test the null hypothesis
that the time series contains no periodicities, that is, has no temporal
correlation. Theiler's algorithm~$0$ generates surrogate time series
having no temporal correlation by simply shuffling the original time
series, or put another way, the surrogates are i.i.d. noise having the
same same rank distribution as the original time series~\cite{jT96a}.

Our proposal is to use optimal RARM as the test for periodicity, that is,
if the optimal RARM is non-trivial in that $k>0$ in~(\ref{rarm}), then
periods are present in the time series. To believe the validity of this
test one must require that if the optimal RARM detects a period in a time
series, then it must not detect any period in algorithm~$0$
surrogates~\cite{jT96a,pivotal_thing}. This surrogate test must be applied
to each data set for which an optimal RARM has been constructed to ensure
that the structure detected in {\em each} data set is genuine. That is, we
propose that an algorithm~$0$ surrogate test is a necessary part of the
procedure of detecting periodicity using an optimal RARM. If RARM methods
identify periodicity in the surrogates then this is clear evidence of
false identification of periodicity in the data. However, if the RARM
algorithm detects no periodicity in the surrogates then periodicity
identified in the original data is genuine.  To ensure the reverse
implication (\ref{onlyifstate}) holds one need only apply an algorithm $0$
surrogate calculation.

\section{Calculations}
\label{calculations}
In this section we demonstrate with artificial and experimental data that
RARM detects periodic behavior (\ref{ifstate}) if and (\ref{onlyifstate})
only if it is present in the original time series. To demonstrate that
RARM detects periodic behavior {\em if} it is present in the data we
construct artificial data contaminated with noise and demonstrate the
effectiveness of the RARM algorithm. We compare the RARM results to
traditional Fourier spectral and autocorrelation techniques. We repeat
these calculations for some experimental data comparing the RARM algorithm
and traditional techniques.  To demonstrate that our RARM algorithm
detects periodic behavior {\em only if} it is present in the data we apply
the method of surrogate data.

In section \ref{babies} we describe the application of these techniques to
detect periodicities in recordings of infant respiratory patterns during
natural sleep. Section \ref{artificial} applies these methods to
artificial data sets to demonstrate the effectiveness of these techniques
compared to traditional methods.  Section \ref{exp_res} describes the
application of these same methods to global climatic data.

\subsection{Infant respiratory data}
\label{babies}

\begin{figure*}
  \epsfxsize 145mm
%\[\epsfbox{../pictures/rarm/Lochlan1_tide.eps}\]
\[\epsfbox{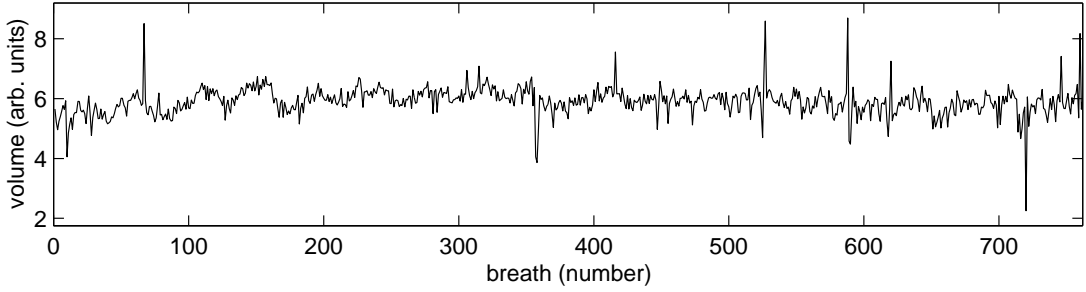}\]
\caption{{\bf Tidal volume:} The
  horizontal axis is breath number --- each datum in this time series
  corresponds to a single breath. The vertical axis is derived from the
  output from the analogue to digital converter (proportional to
  cross-sectional area measured by inductance plethysmography, arbitrary
  units). For each breath the minimum and maximum value over that breath
  were calculated and the difference recorded. This data set consists of
  $762$ points recorded from a $21$ week old male during $24$ minutes of
  continuous stage $2$ sleep. This study had approval from the ethics
  committee of Princess Margaret Hospital.  The parents of this subject
  were informed of the procedure, and its purpose, and had given consent.
  The recording took place during a scheduled overnight sleep study at
  Princess Margaret Hospital.}
\label{data-exp}
\end{figure*}

Using inductance plethysmography we have collected measurements of
cross-sectional area of the abdomen of infants during natural sleep. From
these measurements we extract a measure that can be related to the breath
volume \cite{dimension_thing}. Figure \ref{data-exp} gives an example of
data collected in this way.

\begin{figure*}
  \epsfxsize 145mm
%\[\epsfbox{../pictures/rarm/data-conventional.eps}\]
\[\epsfbox{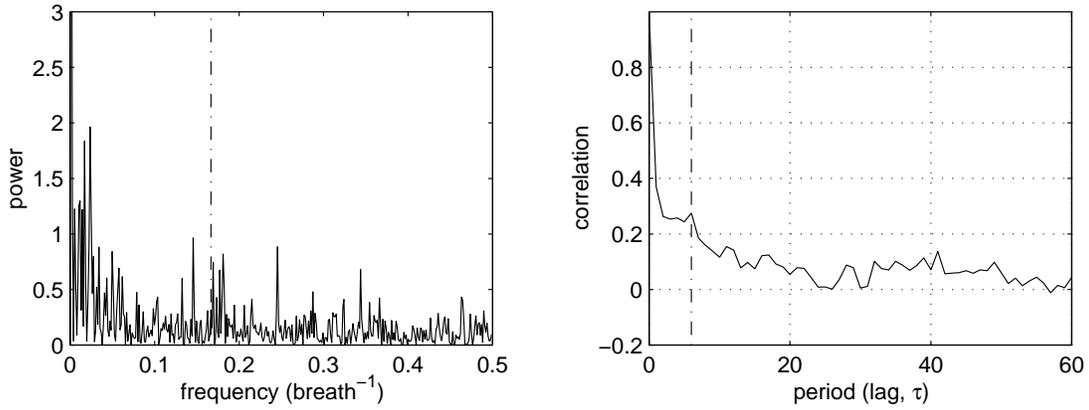}\]
\caption{{\bf Spectral techniques:}  Estimates of the power spectrum
  (arbitrary units) and autocorrelation function for the data illustrated
  in figure \ref{data-exp}. The RARM detected periodic motion over a
  period of $6$ data points, see equation (\ref{loch_model}). A vertical
  dot-dashed line marks the location of period 6 behavior in both the
  frequency (power spectrum) and time (autocorrelation) domain. A peak in
  the autocorrelation function corresponds exactly with the period $6$
  behavior detected by RARM. The power spectrum has a peak close to a
  frequency of $6^{-1}\approx 0.166667$. A period of $6$ is the closest
  integer value to the peak evident at this location in the power
  spectrum. Whilst both power spectra and autocorrelation detect behavior
  with a period of $6$ these results are not as conclusive as the RARM
  algorithm.}
\label{exp-calc}
\end{figure*}

We applied our RARM procedure to the data illustrated in figure
\ref{data-exp} and obtained a model of the form
\begin{equation}
y_t=a_0+a_1y_{t-1}+a_2y_{t-6}+e_t
\label{loch_model}
\end{equation}
where $a_0\approx 2.945206$, $a_1\approx 0.300739$ and $a_2\approx
0.202056$. Figure \ref{exp-calc} shows the result of analysis of this data
set with a fast Fourier transform algorithm (MATLAB's {\tt spectrum}
command.) and an estimate of the autocorrelation function.  Both these
techniques yield small peaks at the same value (that is, $6$) and are
consistent with the results of our RARM algorithm. However, the results
are not as unambiguous as the results of the RARM algorithm. That is, the
RARM detects a periodicity that is not strong enough to be unambiguously
identified by spectral methods.

For many time series of breath size \cite{experiments} we have computed
autocorrelation and Fourier spectral estimates. We have applied our RARM
algorithm to each data set and compared this to the result of applying
traditional techniques. For these data the period of periodic behavior
detected by the RARM algorithm is consistent with the periods detected by
autocorrelation. That is, if RARM detects periodic behavior, then it is of
the same period as that detected by the autocorrelation estimate (if the
autocorrelation detects periodic behavior).  Furthermore, if RARM does not
detect periodic behavior, then neither does the autocorrelation estimate.
The traditional techniques will often fail to detect periodic behavior
when the RARM algorithm does detect it.

We have provided experimental evidence that the RARM technique detects
periodic behavior when it does occur. Now we will demonstrate that the
RARM technique does not lead to spurious identification of periodic
behavior. That is, we will show that if the RARM algorithm detects
periodic behavior, {\em then} there is periodic behavior in the data. To
do this we apply a surrogate data algorithm which will ensure that false
indications of periodicities can always be identified.

For the data illustrated in figure \ref{data-exp}, none of $100$
surrogates generated by shuffling the data exhibited periodic behavior of
any period. This calculation was repeated with another $48$ data sets
\cite{experiments}. In all $49$ cases the RARM failed to detect periodic
behavior in the surrogate data in at least $99$ (of $100$) surrogates of
each data set. This indicates that the RARM algorithm does not identify
periodicities not present in the data.

\subsection{Artificial data}
\label{artificial}

\begin{figure*}
  \epsfxsize 145mm
%\[\epsfbox{../pictures/rarm/arnoise.eps}\]
\[\epsfbox{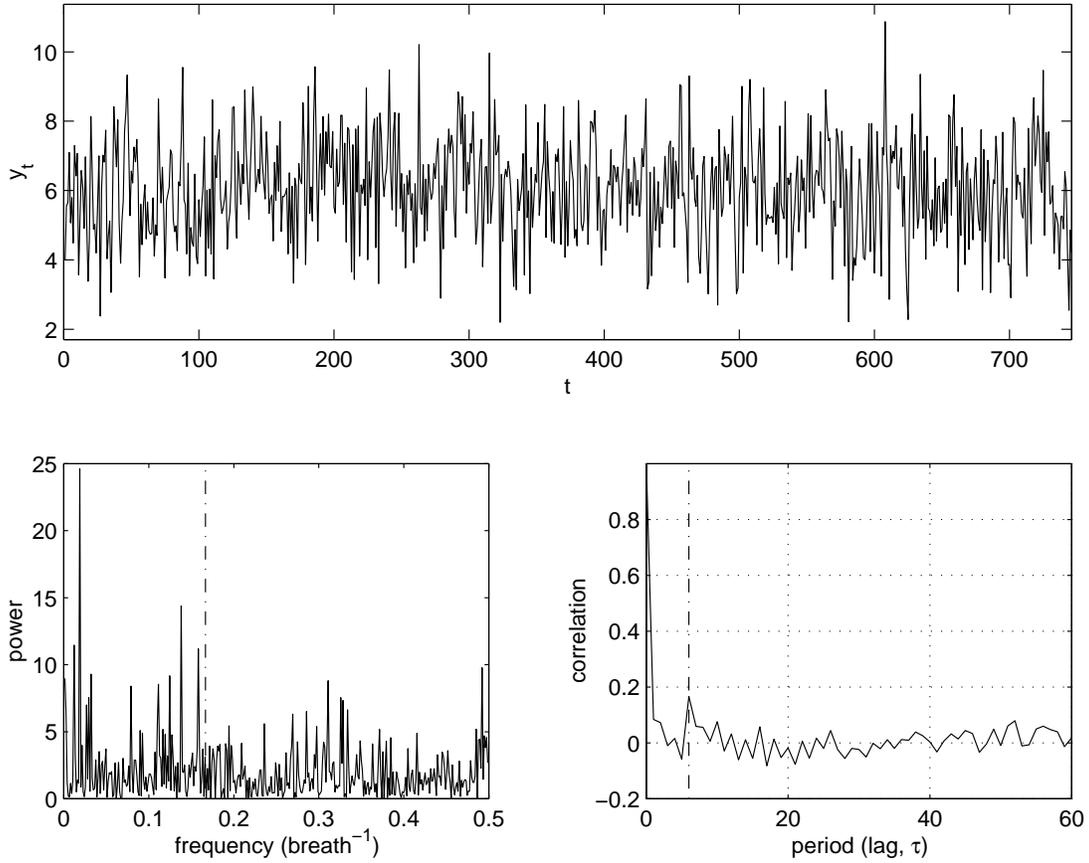}\]
\caption{{\bf Artificial data:} A data set of $764$ realization of the 
  process described by (\ref{artrarm}) with normal observational noise,
  standard deviation 1. This linear model is of the same form as that
  predicted from the model of the data in figure \ref{data-exp}.  Also
  shown in the power spectrum (arbitrary units) and autocorrelation
  estimate for this data set. For this data set RARM gave a clear
  indication of period $6$ behavior. The dot dashed line on the power
  spectrum and autocorrelation function corresponds to the period of $6$
  detected by RARM.}
\label{art}
\end{figure*}

\begin{figure*}
  \epsfxsize 145mm
%\[\epsfbox{../pictures/rarm/arlong.eps}\]
\[\epsfbox{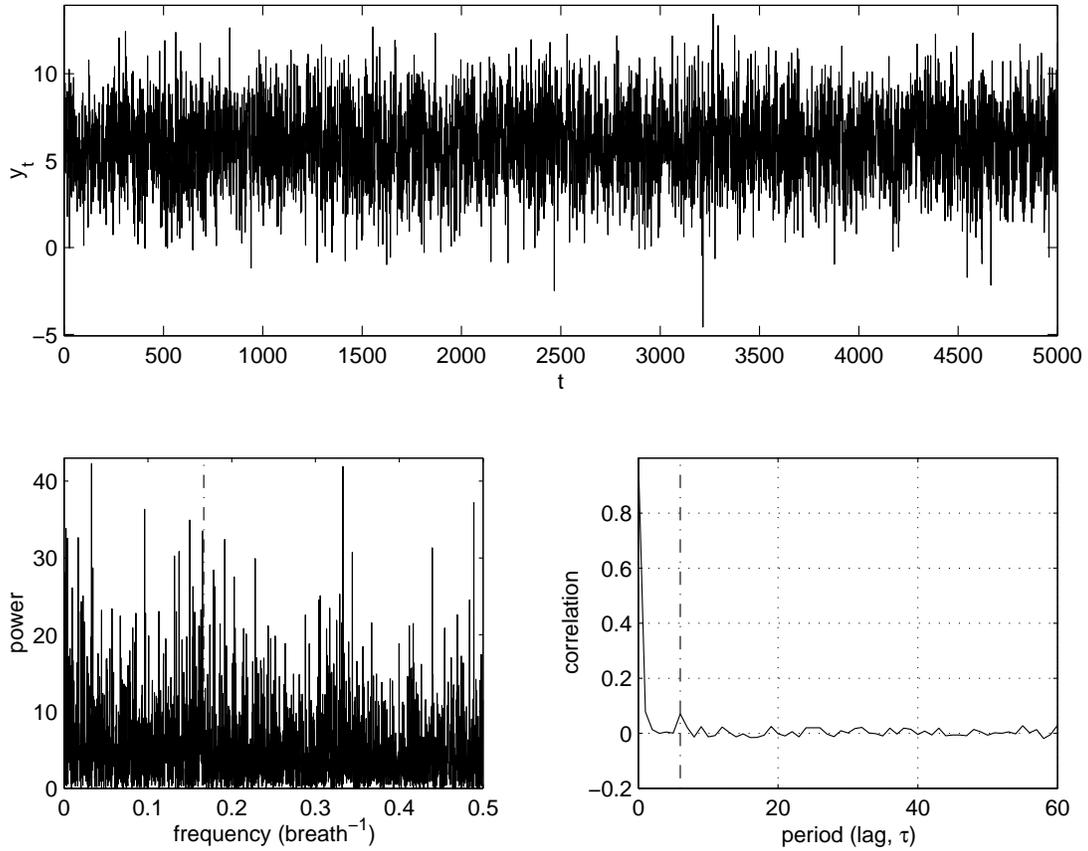}\]
\caption{{\bf Artificial data:} Data from an reduced autoregressive of the
  same form as that predicted from the model of the data in figure
  \ref{data-exp}. This data sets consists of $5000$ realizations of
  (\ref{artrarm}) with observational noise, standard deviation 2. Also
  shown in the power spectrum (arbitrary units) and autocorrelation
  estimate for this data set. For this data set RARM gave a clear
  indication of period $6$ behavior. The dot dashed line on the power
  spectrum and autocorrelation function corresponds to the period of $6$
  detected by RARM.}
\label{longart}
\end{figure*}

In this section we use the optimal RARM from section \ref{babies} as a
basis for generating noisy artificial data with a known periodicity.  From
(\ref{loch_model}) we use the model
\begin{equation}
y_t=a_0+a_1y_{t-1}+a_2y_{t-6}+e_t
\label{artrarm}\end{equation}
(where $a_0\approx 2.945206$, $a_1\approx 0.300739$ and $a_2\approx
0.202056$, as above) to generate an artificial data set $y$.  To this data
we add observational noise $\epsilon_t$ and apply the above analysis to
the series $z$, $z_t=y_t+\epsilon_t$. Figure \ref{art} demonstrates the
result of this technique for an artificial data set of the same length as
the data and normal observational noise with standard deviation $1$
($e_t,\epsilon_t\sim N(0,1)$). Figure \ref{longart} is the result of the
same technique for a longer data set ($5000$ data points) and more
observational noise ($e_t\sim N(0,1)$ and $\epsilon_t\sim N(0,2)$). In
both cases RARM clearly identified periodic behavior with period $6$. For
the time series in shown in figures \ref{art} and \ref{longart} we
constructed $100$ algorithm $0$ surrogates. none of them exhibited
periodicity detected by RARM.

The traditional Fourier spectral and autocorrelation techniques identify
the same period as the RARM technique for the shorter, but less noisy data
illustrated in figure \ref{art}. However, for the data shown in figure
\ref{longart} the RARM technique has identified periodicities that are not
obvious from traditional techniques. Furthermore, it should be noted that
in all cases that the results of the autocorrelation and spectral methods
are not clear cut. For reasonably long, but extremely noisy data sets the
RARM algorithm still provides a decisive and accurate estimate of the
period of periodic behavior present in data.
 
\subsection{Global climatic data}
\label{exp_res}

\begin{figure*}
  \epsfxsize 145mm
%\[\epsfbox{../pictures/rarm/gltemp.eps}\]
\[\epsfbox{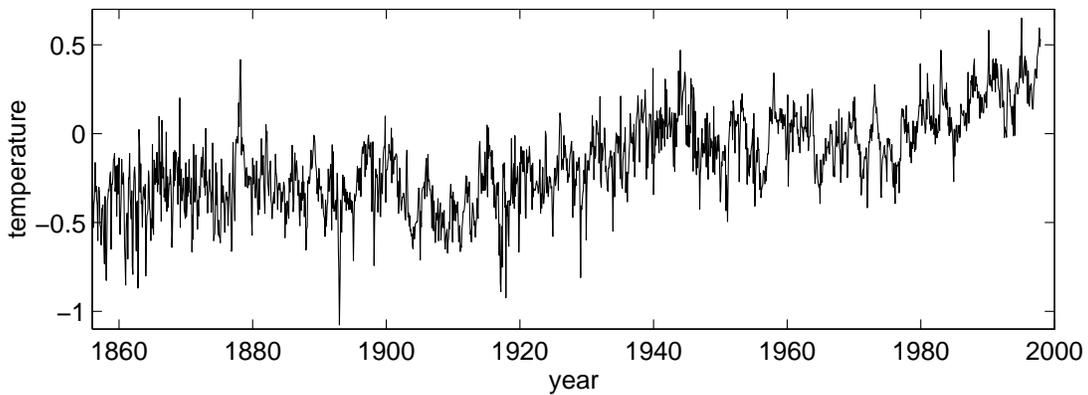}\]
\caption{{\bf Global air temperature:} Monthly global air temperature
  measured as deviation (in degrees Celsius) from monthly mean temperature
  for the period 1856-1997 ($1704$ data).}
\label{gltemp}
\end{figure*}

In this section we describe the application of these techniques with noisy
physical data. The time series we use here is monthly deviations from
monthly mean global air temperatures over the period 1856--1997
\cite{weblink}. These global air temperature measurements are obtained by
averaging observations at many spatially separated sites on the globe.
Figure \ref{gltemp} shows the complete data set. A more detailed
discussion of this data may be found in \cite{nN96}.  Analysis using the
methods described in this paper demonstrates the presence of periodic
fluctuation over periods of $7$ months, $2$ years and $45$ months\cite{elnino}. Fourier
spectral and autocorrelation estimates were also applied (after
de-trending this time series) and the results are illustrated in figure
\ref{gltemp}. From $100$ algorithm $0$ surrogates RARM did not detect
periodicity in $99$ of them. These results demonstrate the presence of
genuine periodic fluctuation in this time series and that the fluctuation
is difficult to detect with traditional techniques. An advantage of the
RARM technique is that no de-trending is required. The results of the RARM
algorithm are not effected by trends or non-stationarity.

\section{Conclusions}

We have provided theoretical and experimental evidence to support the use
of RARM techniques to detect periodic behavior in noisy experimental time
series.  The concept of minimum description length ensures that a RARM
built with an MDL modeling criterion will detect any periodicities present
in the data. We provided numerical evidence using experimental and
artificial data to support this. Moreover these calculations have
demonstrated that the RARM algorithm provides an accurate and decisive
method of detecting periodicities that is more sensitive than Fourier
spectrum or autocorrelation methods.

By applying surrogate data techniques we have demonstrated that the RARM
algorithm did not identify periodicities in temporally uncorrelated
surrogates. This is strong experimental evidence that the RARM algorithm
is robust against identification of false periodicities. It does not
identify behavior not present in the original system. However this result
has only been supported by numerical evidence and does not imply that true
identification with arbitrary data. To guard against false positives we
recommend application of surrogate data tests, as discussed in this paper.
Periodicity detected using RARM are genuine provided RARM detects no
periodicity in i.i.d. surrogates.

\section*{Acknowledgments}

We wish to thank Madeleine Lowe and Stephen Stick of Princess Margaret
Hospital for Children for supplying the infant respiratory data, and for
physiological guidance. We also thank Tiempo Climatic Cyberlibrary for
making the global climatic data used in this article easily available.

\bibliographystyle{prsty} %\bibliography{../bibliography,rarm_notes}

\begin{thebibliography}{10}
\bibitem{elnino}
These three periods at $7$, $24$, and $45$ months are likely to be related to
  (respectively) a seasonal cycle, the quasi-biennial cycle, and an El Nino
  effect (Private Communication: Mick Kelly, Climatic Research Unit, University
  of East Anglia).
\end{thebibliography}

\newpage
\onecolumn

\begin{figure*}
   \epsfxsize 145mm
%\[\epsfbox{../pictures/rarm/gltemp-calc.eps}\]
\[\epsfbox{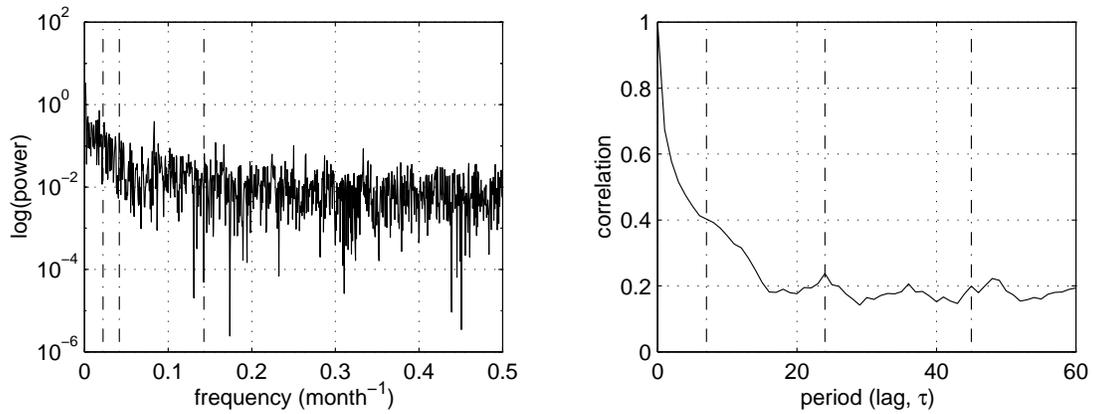}\]
\caption{{\bf Spectral techniques:}  Estimates of the power spectrum
  and autocorrelation function for the data illustrated in figure
  \ref{gltemp}. The data in figure \ref{gltemp} was linearly de-trended
  before calculating Fourier spectrum and autocorrelation. The RARM
  detected periodic motion over a period of $7$, $24$ and $45$ months. A
  vertical dot-dashed line marks the location of period $7$, $24$ and $45$
  behavior in both the frequency and time domain. A peak in the
  autocorrelation function corresponds exactly with the period $24$ and
  $45$ behavior detected by RARM. The power spectrum has a peak close to a
  frequency of $45^{-1}\approx 0.0222$. Whilst both power spectra and
  autocorrelation detect behavior with a period of $24$ and $45$ these
  results are not as conclusive as the RARM algorithm.}
\label{gltemp-calc}
\end{figure*}

\end{document}